\documentclass[aps,pre,twocolumn,superscriptaddress,longbibliography,nobibnotes,nodoi,nourl,noeprint]{revtex4-2}
\usepackage{amsmath,amssymb,physics,bm,siunitx}
\usepackage{xcolor,graphicx,soul}
\usepackage[colorlinks=true,citecolor=blue,urlcolor=blue,linkcolor=black]{hyperref}
\usepackage{subfiles}
\usepackage{setspace}

\def\rms{\rm\scriptscriptstyle}
\def\dd{{\rm d}}

\begin{document}
	
\title{Scaling laws for single-file diffusion of adhesive particles}

\author{S\"{o}ren~Schweers}
\email{sschweers@uos.de}
\affiliation{Universit\"{a}t Osnabr\"{u}ck, Fachbereich Physik, Barbarastra{\ss}e 7, D-49076 Osnabr\"uck, Germany}

\author{Alexander P.\ Antonov}
\email{alantonov@uos.de}
\affiliation{Universit\"{a}t Osnabr\"{u}ck, Fachbereich Physik, Barbarastra{\ss}e 7, D-49076 Osnabr\"uck, Germany}

\author{Artem Ryabov}
\email{rjabov.a@gmail.com}
\affiliation{Charles University, Faculty of Mathematics and Physics, Department of Macromolecular Physics, V Hole\v{s}ovi\v{c}k\'{a}ch 2, 
CZ-18000 Praha 8, Czech Republic}

\author{Philipp Maass} 
\email{maass@uos.de}
\affiliation{Universit\"{a}t Osnabr\"{u}ck, Fachbereich Physik, Barbarastra{\ss}e 7, D-49076 Osnabr\"uck, Germany}

\date{March 1, 2023} 

\begin{abstract}
Single-file diffusion refers to the
Brownian motion in narrow channels where particles cannot pass each other. 
In such processes, the diffusion of a tagged particle
is typically normal at short times and becomes subdiffusive at long times. 
For hard-sphere interparticle interaction, 
the time-dependent mean squared displacement of a tracer is well understood.
Here we develop a scaling theory for adhesive particles.
It provides a full description of the time-dependent diffusive behavior with a scaling function that depends on an
effective strength of adhesive interaction. Particle clustering induced by the adhesive interaction slows down the diffusion at short times, while it enhances subdiffusion at long times. The enhancement effect can be quantified in measurements irrespective of how tagged particles are injected into the system. Combined effects of pore structure and particle adhesiveness should 
speed up translocation of molecules through narrow pores.
\end{abstract}

\maketitle

Diffusive motion in narrow pores is an important process in many biological, chemical and engineering systems. 
When particles in these confined structures cannot overtake each other, the process 
is referred to as single-file diffusion \cite{Kaerger:2014, Nygard:2017, Taloni/etal:2017, Bukowski/etal:2021}.
Prominent examples are the tracer diffusion 
in zeolites \cite{Hahn/etal:1996, Hahn/Kaerger:1998, Chmelik/etal:2018}, in colloidal systems \cite{Wei/etal:2000, Cui/etal:2002, Lin/etal:2002,  Lutz/etal:2004a, Lutz/etal:2004b, Lin/etal:2005, Koeppl/etal:2006, Henseler/etal:2010},
nanotubes \cite{Cheng/Bowers:2007, Das/etal:2010, Dvoyashkin/etal:2014, Cao/etal:2018, Zeng/etal:2018}, 
in membrane channels and pores \cite{Bauer/Nadler:2006, Kahms/etal:2009, Yang/etal:2010,  Luan/Zhou:2018, Zhao/etal:2018, Kaerger/etal:2021}, 
and in a macroscopic system
of electrically interacting metallic beads confined to a ring \cite{Coste/etal:2010}.
In the long-time limit the mean squared displacement $\langle\Delta x^2(t)\rangle$ of a tagged particle in single-file diffusion
does not grow linearly but with the square root of time $t$ \cite{Harris:1965, Levitt:1973, Arratia:1983, Hahn/Kaerger:1996, Kollmann:2003, Lizana/Ambjornsson:2008, Dvoyashkin/etal:2013, Ryabov:2016, Dolai/etal:2020, Wittmann/etal:2021, Banerjee/etal:2022},
\begin{equation}
\langle \Delta x^2(t)\rangle\sim 2D_{1/2} \sqrt{t}\,,\hspace{1em}\mbox{for}\hspace{0.3em}t\to\infty.
\label{eq:subdiffusion}
\end{equation}
The coefficient $D_{1/2}$ quantifies the speed of the spreading similarly as the diffusion coefficient in normal diffusion.

Generally, the mean squared displacement shows a crossover from a normal
diffusive behavior $\langle \Delta x^2(t)\rangle\sim 2Dt$ at short times to the subdiffusive law \eqref{eq:subdiffusion}
at long times. In simple systems, one could expect the crossover time $t_\times$ between 
the normal and subdiffusive regime to be determined by the condition that the root of the mean squared displacement
equals the mean distance $1/\rho$ between particles, where $\rho$ is the particle number density. 
This gives $2Dt_\times\sim 1/\rho^2$, and by matching $2Dt_\times$ with
$2D_{1/2} t_\times^{1/2}$, one obtains $D_{1/2}\sim \sqrt{D}/\rho$. In lattice systems at low densities, or continuous space systems 
of point particles, this relation between $D$ and $D_{1/2}$ is indeed often obtained \cite{vanBeijeren/etal:1983, Arratia:1983, Lizana/Ambjornsson:2008, Leibovich/Barkai:2013}.

Less is known about the behavior of the mean squared displacement in single-file systems of particles 
with attractive and repulsive interactions. A prominent model  to capture major characteristics of real interactions with a  repulsive core and adhesive part is the model of sticky hard spheres \cite{Baxter:1968, Percus:1982}, where the
pair interaction $V(r)$ between particles is given by 
\begin{equation}
\exp[-V(r)/k_{\rms B} T] = \Theta(r-\sigma)+\gamma\sigma\delta_+(r-\sigma)\,.
\label{eq:pair-interaction}
\end{equation}
Here, $\sigma$ is the particle diameter, $k_{\rms B} T$ is the thermal energy, $\Theta(.)$
is the Heaviside step function [$\Theta(x)=1$ for $x>0$ and zero otherwise], and $\delta_+(r)$ is the right-sided 
$\delta$-function, i.e.\ for any test function $h(r)$ and $\epsilon>0$, 
it holds $\int_0^\epsilon\dd r\, h(r)\delta_+(r)=h(0)$. 
The  Heaviside function $\Theta(r-\sigma)$ in Eq.~\eqref{eq:pair-interaction}
implies that $V(r)$ is infinite for $r<\sigma$, i.e.\ it takes into account the
hardcore repulsion. The function $\gamma\sigma\delta_+(r-\sigma)$ describes an additional
attractive contact interaction, where $\gamma\sigma$ quantifies the adhesive strength. 
We refer to the dimensionless parameter $\gamma$ as the stickiness.

In this Letter we show that the mean squared displacement of sticky hard spheres in single-file systems can be described by scaling laws, where the scaling function depends on an effective 
stickiness parameter between the spheres, which combines $\gamma$ 
with the particle number density $\rho$. Exact results are given for 
the limiting behavior  of the scaling function at short and long scaled times. The coefficient $D_{1/2}$ in
Eq.~\eqref{eq:subdiffusion}
is proportional to the square root of the isothermal 
compressibility of the system, as can be inferred from a general result for the long-time asymptotics \cite{Kollmann:2003}, including
the case of underdamped Brownian motion \cite{Delfau/etal:2011}. 
The dependence on the compressibility implies that subdiffusion becomes faster for attractive and slower for repulsive interactions. This is in 
contrast to what is typically found in normal diffusion and different from what has been seen for single-file diffusion in some one-dimensional lattice models with attractive interactions \cite{Centres/Bustingorry:2010, Fouad/Noel:2017}. 

The overdamped Brownian motion of the particles with positions $x_i$, $i=1,\ldots,N$
is described by the Langevin equations
\begin{equation}
\frac{\dd x_i}{\dd t}=\sqrt{2D}\,\xi_i(t),
\label{eq:langevin}
\end{equation}
where $D$ is the bare diffusion coefficient and $\xi_i(t)$ are Gaussian white noise processes with 
zero mean and correlation functions $\langle \xi_i(t) \xi_j(t') \rangle = \delta_{ij}\delta(t - t')$. 
The treatment of the
hardcore and adhesive interactions needs special care due to their singular nature. We have applied our recently developed Brownian cluster dynamics algorithm to tackle this problem \cite{Antonov/etal:2022c}.
For determining equilibrium properties, we have performed 
also Monte Carlo simulations based on the method developed in Refs.~\cite{Miller/Frenkel:2004_1, Miller/Frenkel:2004_2}. 

\begin{figure}[t!]
\includegraphics[width=\columnwidth]{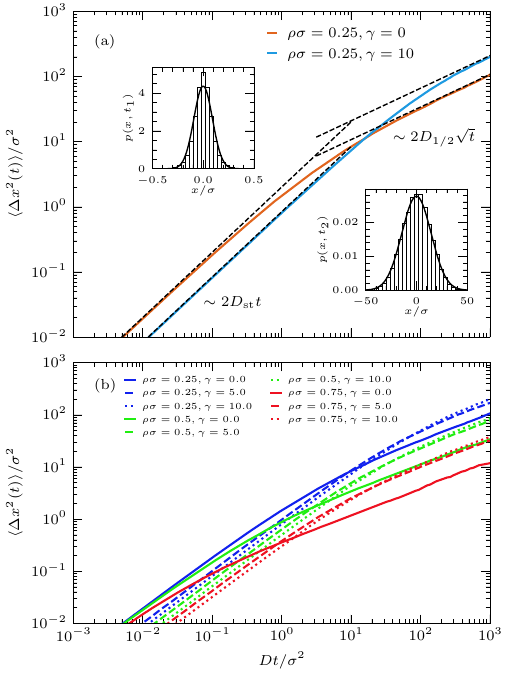}
\caption{(a) Mean squared displacement of a tagged particle for scaled density (coverage) $\rho\sigma=0.25$ in the absence of adhesive interactions ($\gamma=0$) and for stickiness $\gamma=10$. The insets show the diffusion propagator at fixed times in the short-time regime ($Dt_1/\sigma^2=10^{-2}$) and the long-time regime ($Dt_2/\sigma^2=10^3$) for $\gamma=10$. 
(b) Mean squared displacement for various $\rho\sigma$ and $\gamma$.}
\label{fig:mean_squared_displacement}
\end{figure}

Figure~\ref{fig:mean_squared_displacement}(a) shows simulation results for $\langle \Delta x^2(t)\rangle$ at one number density 
$\rho=1/4\sigma$
for strong adhesive interaction $\gamma=10$ and for $\gamma=0$  \footnote{The simulations have been performed under periodic boundary conditions for typical system sizes of $800\sigma$. To obtain good numerical accuracy, averages were taken over 250 equilibrated initial particle configurations. At stickiness $\gamma=10$, the CPU time for a simulation run of one configuration was 5h for $\rho\sigma=0.25$ and 45h for $\rho\sigma=0.8$ on one core of an AMD EPYC 7742 processor.}.
The initial distribution of the tagged particle is the equilibrium one. In an equilibrium configuration,
the tagged particle can be part of different clusters of particles in contact. A sequence of particles with positions
$x_i=x_1+(i-1)\sigma$, $i=1,\ldots,n$ and no particles at positions $x_1-\sigma$ and $x_n+\sigma$ forms a cluster
of size $n$ ($n$-cluster); single particles are 1-clusters. 

At short times, the motion of clusters is not affected by the presence of neighboring clusters and $\langle \Delta x^2(t)\rangle$ grows linearly 
in time, corresponding to normal diffusion. At long times, the mean square displacement 
shows the subdiffusion characteristic for single-file Brownian motion,
$\langle \Delta x^2(t)\rangle\sim t^{1/2}$. 
In the insets of Fig.~\ref{fig:mean_squared_displacement}(a), we display the diffusion propagator 
$p(x,t)$ at two fixed times in the short- and long-time regime. 
For all times, this propagator is a Gaussian function, $p(x,t)\propto\exp[-x^2/2\langle \Delta x^2(t)\rangle]$.
Note that $\langle \Delta x^2(t)\rangle$
at long times is enhanced by the adhesive interaction. This speed-up of subdiffusion may be 
unexpected at first sight because attractive particle interactions usually slow down Brownian motion.

Figure~\ref{fig:mean_squared_displacement}(b) shows simulation results for various $\rho\sigma$ and $\gamma$. The speed-up effect is always present, which can be seen by comparing the curves for different $\gamma$ (same color) at fixed $\rho\sigma$ values
(solid, dashed or dotted lines). The change of functional behavior of 
$\langle \Delta x^2(t)\rangle$ with $\rho\sigma$ and $\gamma$ seems to be complicated.
 
We now develop a scaling theory that fully describes the behavior of $\langle \Delta x^2(t)\rangle$ in this many-body system. We start by determining the behavior in the short-time limit $t\to0$, which can be inferred from the distribution of cluster sizes in the equilibrium state. 
To derive the cluster size distribution, we can build on 
exact results for thermodynamic and 
structural properties of the sticky-core fluid in equilibrium
\cite{Salsburg/etal:1953, Tago/Katsura:1975, Yuste/Santos:1993}. 
For the pair correlation function, it was found \cite{Yuste/Santos:1993}
\begin{equation}
g(r)=\frac{1}{\rho}\sum_{n=1}^\infty\Biggl\{
q^n\delta_+(r-n\sigma)
+f_n(r-n\sigma)\Theta(r-n\sigma)\Biggr\}\,,
\label{eq:g(r)}
\end{equation}
where $f_n(r)$ are smooth functions of $r$, and
\begin{equation}
q=\frac{\sqrt{1+4\tilde\gamma}-1}
{\sqrt{1+4\tilde\gamma}+1}\,,
\end{equation}
with
\begin{equation}
\tilde\gamma=\frac{\gamma\rho\sigma}{1-\rho\sigma}\,.
\end{equation}
This dimensionsless parameter is the stickiness multiplied by the ratio of particle diameter $\sigma$ to the
mean size $1/\rho-\sigma$ of free space between particles. It describes the effective stickiness.

As shown in the supplemental material (SM),
the $\delta$-functions in Eq.~\eqref{eq:g(r)} with the amplitudes $q^n/\rho$ 
follow if the cluster size distribution is the geometric distribution 
\begin{equation}
w_n=(1-q)q^{n-1}\,.
\label{eq:size-distribution}
\end{equation}
The mean cluster size then is given by
\begin{equation}
\bar n(\tilde\gamma)=\frac{1}{1-q}=\frac{1}{2}\left(\sqrt{1+4\tilde\gamma}+1\right)\,.
\label{eq:mean-size}
\end{equation}
Both Eqs.~\eqref{eq:size-distribution} and \eqref{eq:mean-size} agree with the simulated data, see SM.

Initially, the tagged particle is part of an $n$-cluster 
with probability $\propto nw_n$, and the center of mass of this $n$-cluster has a diffusion constant $D/n$.
Accordingly, the short-time diffusion coefficient $D_{\rm st}$ of the tagged particle is obtained by averaging $D/n$
over the distribution $\propto nw_n$, yielding
\begin{equation}
D_{\rm st}(\tilde\gamma,\rho\sigma)=\frac{D}{\bar n(\tilde\gamma)}
=\frac{2D}{\sqrt{1+\dfrac{4\gamma\rho\sigma}{1-\rho\sigma}}+1}\,.
\label{eq:Dst}
\end{equation}
When time increases, the Brownian motion of the tagged particle becomes mitigated due to the hindrance of free diffusion by neighboring particles, leading to subdiffusion as described by Eq.~\eqref{eq:subdiffusion}.

For determining the dependence of $D_{1/2}(\tilde\gamma,\rho\sigma)$ 
on $\tilde\gamma$ and $\rho\sigma$, let us first consider a situation, where
fragmentations and mergers of the initial clusters during the course of time are neglected. This situation corresponds to a 
diffusion of a tagged particle in a random mixture of clusters with different fixed sizes given by the 
geometric distribution \eqref{eq:size-distribution}.

To derive the diffusion coefficient of a tagged particle in this cluster mixture, we use scaling arguments. If
all clusters are represented by ones having the same mean size $\bar n$, 
the crossover from the  normal diffusive behavior at short times to the subdiffusive one at long times will occur 
when the root mean square displacement becomes
proportional to the average spacing between the clusters.
Accordingly, the root of the mean squared displacement at the crossover time $t_\times$ should be proportional to 
$\bar n(1/\rho-\sigma)$. This implies
$2D_{\rm st}\,t_\times\sim [\bar n(\tilde\gamma)\,(1/\rho-\sigma)]^2$, yielding
\begin{equation}
t_\times(\tilde\gamma,\rho\sigma)\sim
\frac{[\bar n(\tilde\gamma)(1/\rho-\sigma)]^2}{2D_{\rm st}(\tilde\gamma,\rho\sigma)}
=\frac{\bar n(\tilde\gamma)^3(1-\rho\sigma)^2}{2D\rho^2}\,.
\label{eq:tx}
\end{equation}
Knowing $t_\times$, we expect the mean squared displacement to obey the following scaling behavior,
\begin{equation}
\langle\Delta x^2(t)\rangle
=\bar n(\tilde\gamma)^2\,\frac{(1-\rho\sigma)^2}{\rho^2}\,F\left(\frac{t}{t_\times(\tilde\gamma,\rho\sigma)}\right)\,.
\label{eq:msd-scaling-fixed-cluster-size}
\end{equation}
Here, $F(u)$ is a scaling function with $F(u)\sim u$ for $u\to0$ and $F(u)\sim F^\infty u^{1/2}$ 
for $u\to\infty$, where $F^\infty$ is a constant.
In the limit of hardcore interacting point particles ($\sigma=0$, $\gamma=0$), we obtain $\tilde\gamma=0$ and $\bar 
n(\tilde\gamma=0)=1$, and it is known that $D_{1/2}=\sqrt{D/\pi}/\rho$ for equilibrated configurations \cite{Harris:1965, Levitt:1973, vanBeijeren/etal:1983, Lizana/Ambjornsson:2008, Leibovich/Barkai:2013}. 
Accordingly, it must hold $F^\infty=\sqrt{2/\pi}$. Simulations for the system with fixed cluster sizes confirm the scaling behavior predicted 
by Eq.~\eqref{eq:msd-scaling-fixed-cluster-size}, see SM.

Turning back to the system of adhesive particles, the short-time diffusion coefficient in Eq.~\eqref{eq:Dst} must describe the behavior in the 
limit $t\to0$, but we cannot expect $D_{\rm st}(\tilde\gamma,\rho\sigma)$ to be unaffected by the possible fragmentation and mergers of clusters in 
the whole short-time regime $t\lesssim t_\times$. 

However, our analysis above suggests that mean squared displacements can be scaled along a curve in the $\rho$-$\gamma$ plane, where 
the effective stickiness $\tilde\gamma$ is constant. The scaling
\eqref{eq:msd-scaling-fixed-cluster-size} should
hold for $\langle\Delta x^2(t)\rangle$ along such curves, with a $\tilde\gamma$-dependent 
scaling function $F_{\tilde\gamma}(t/t_\times(\tilde\gamma,\rho\sigma))$. The functional behavior of $F_{\tilde\gamma}(u)$ for
$u\to0$ and $u\to\infty$ must be unchanged, but we have to consider now a $\tilde\gamma$-dependent 
amplitude factor $F^\infty_{\tilde\gamma}$. 

\begin{figure}[t!]
\includegraphics[width=\columnwidth]{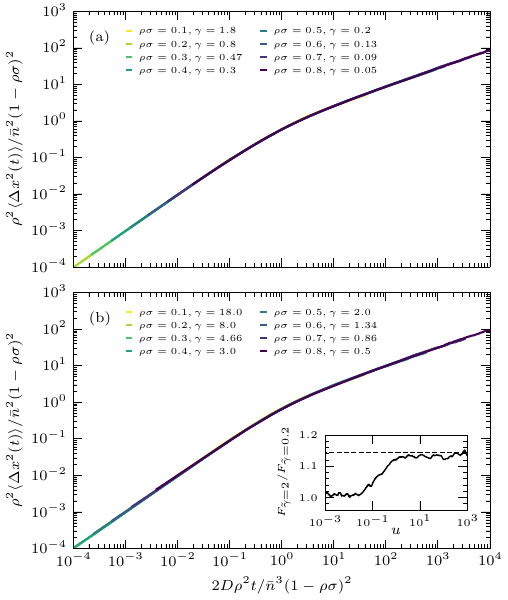}
\caption{Scaled mean squared displacement as a function of scaled time according to 
Eq.~\eqref{eq:msd-scaling-fixed-cluster-size}. 
The data for various densities $\rho$ and adhesive strengths $\gamma$ collapse onto common master curves for
(a) fixed $\tilde\gamma=0.2$ and (b)
fixed $\tilde\gamma=2$.
The inset in (b)
shows the ratio
$F_{\tilde\gamma=2}(u)/F_{\tilde\gamma=0.2}(u)$ of the two scaling functions, which increases from one for $u\to0$
to the constant value 1.144 predicted by Eq.~\eqref{eq:Finf} for $u\to\infty$. The constant is indicated by the dashed horizontal line.}
\label{fig:msd_scaled}
\end{figure}

Figures~\ref{fig:msd_scaled}(a) and (b) show correspondingly 
scaled simulation data of the mean square displacements for various $\rho$ and $\gamma$ at
fixed $\tilde\gamma=0.2$ [Fig.~\ref{fig:msd_scaled}(a)] and $\tilde\gamma=2$ [Fig.~\ref{fig:msd_scaled}(b)].
They are in excellent agreement with the scaling prediction. 
The inset in Fig.~\ref{fig:msd_scaled}(b) shows that the
scaling functions $F_{\tilde\gamma=2}(u)$ and $F_{\tilde\gamma=0.2}(u)$ are equal for small $u$, 
but differ for larger $u$. Within the numerical uncertainties, the ratio $F_{\tilde\gamma=2}(u)/F_{\tilde\gamma=0.2}(u)$ 
increases monotonically from one for small $u$ to a value of about 1.14 for large $u$.

Surprisingly, we can determine the limit $F^\infty_{\tilde\gamma}$ of the scaling function
$F_{\tilde\gamma}(u)$
from equilibrium properties. This is due to the fact that the coefficient $D_{1/2}$ in Eq.~\eqref{eq:subdiffusion} can be expressed by $D$, $\rho$ and the isothermal compressibility $\chi$ \cite{Kollmann:2003, vanBeijeren/etal:1983}. For the adhesive particle system, $\chi$
is known and we obtain (for details, see SM),
\begin{equation}
D_{1/2}(\tilde\gamma,\rho\sigma)=\sqrt{\frac{k_{\rm B}T\chi D}{\pi\rho}}=\frac{1-\rho\sigma}{\sqrt{\pi}\rho}\sqrt{[2\bar n(\tilde\gamma)-1]D}\,.
\label{eq:D1/2}
\end{equation}
For long times, $2D_{1/2}t^{1/2}$ must equal the
asymptotic behavior of Eq.~\eqref{eq:msd-scaling-fixed-cluster-size}.
Taking $t_\times(\tilde\gamma,\rho\sigma)$ from Eq.~\eqref{eq:tx} this yields
\begin{equation}
F^\infty_{\tilde\gamma}=\sqrt{\frac{2}{\pi}\left[2-\frac{1}{\bar n(\tilde\gamma)}\right]}\,.
\label{eq:Finf}
\end{equation}
This function is weakly increasing from the point particle limit $F^\infty_{\tilde\gamma=0}=\sqrt{2/\pi}$ to 
$F^\infty_{\tilde\gamma=\infty}=\sqrt{2}F^\infty_{\tilde\gamma=0}$ for strong adhesive interaction or, more precisely, strong effective stickiness $\tilde\gamma$.
In agreement with Eq.~\eqref{eq:Finf}, the simulated data of $F_{\tilde\gamma=2}(u)/F_{\tilde\gamma=0.2}(u)$ for large $u$ 
approach the limit  $F^\infty_{\tilde\gamma=2}/F^\infty_{\tilde\gamma=0.2}\cong1.144$, see the
inset of Fig.~\ref{fig:msd_scaled}(b).

In summary we have developed a scaling theory for the time-dependent mean squared displacement of a tagged particle
in single-file diffusion of adhesive hard spheres. The scaling applies to curves in the plane of stickiness and particle densities, where the effective stickiness is constant. The effective stickiness is
the product of stickiness and the ratio of particle diameter to mean size of free space between particles. 
Due to the adhesion, particles gather in clusters with sizes distributed according to a geometric distribution, where the mean cluster size increases with the effective stickiness. The clustering 
slows down the normal diffusive behavior at short times, i.e.\ the short-time diffusion 
coefficient decreases with increasing adhesive strength. By contrast, the subdiffusion at long-times is speeded up with 
raising adhesive strength, because collective density fluctuations are decisive for the tracer to propagate over long distances. These density fluctuations become stronger if the particles form clusters and the mean free space between clusters becomes larger. The crossover from the normal to the subdiffusive regime occurs at a time, which increases with the third power of the mean cluster size and is thus
growing with the adhesiveness.

The effect of the density fluctuations on the coefficient $D_{1/2}$ characterizing the speed of the subdiffusive spreading can be
fully taken into account by considering their long-wavelength behavior, which is given by the zero-wavenumber limit 
of the static structure factor $S(0)$ \cite{Kollmann:2003}. The latter is related to the 
isothermal compressibility $\chi$, $S(0)=k_{\rm B}T\rho\,\chi$.
For hardcore interacting particles, the important role of the compressibility has been pointed out already in early studies of single-file diffusion 
\cite{vanBeijeren/etal:1983}.

An interesting aspect of the dependence of $D_{1/2}$ on $S(0)$ concerns the impact of initial conditions: what matters for $D_{1/2}$ is the initial arrangement of the 
particles in the system, while the initial placement of the tagged particle is irrelevant. In our simulations of sticky hard spheres, 
we show this independence of $D_{1/2}$ on the initial starting position of the tagged particle in the SM. For measurements this implies 
that particles injected as, e.g., radioactive or fluorescenct-labeled tracers need not to be equilibrated with the surrounding structure. 

Furthermore, the sensitivity of $D_{1/2}$ to $S(0)$ can be important to understand different speeds of subdiffusion, as reflected, for example, 
in traversal times of particles through pores. One may also think of utilizing this effect: a patterning of a channel leading to a larger particle 
separation or a clustering of particles should facilitate the diffusive motion through the channel. Systematic investigations of such 
pattern-induced enhancement of single-file diffusion and the effects of stickiness should be possible, e.g., in microfluidic 
devices \cite{Misiunas/Keyser:2019, Driscoll/Delmotte:2019}. For adhesive particles, it is in particular interesting to study also the short-time regime, where the 
adhesiveness and particle clustering can be determined from the short-time diffusion coefficient.

\begin{acknowledgments}
We thank the Czech Science Foundation (Project No.\ 20-24748J) and the Deutsche 
Forschungsgemeinschaft (Project No.\ 432123484) for financial support.
We acknowledge use of a high-performance computing cluster funded by the Deutsche Forschungsgemeinschaft
(Project No.\ 456666331).
\end{acknowledgments}
 

%

\onecolumngrid
\newpage
\renewcommand{\theequation}{S\arabic{equation}}
\renewcommand{\thefigure}{S\arabic{figure}}
\setcounter{equation}{0}
\setcounter{figure}{0}

\begin{center}
\setcounter{page}{1}
{\large\bf Supplemental Material for}\\[2ex]
{\large\bf Scaling laws for single-file diffusion of adhesive particles}\\[2ex]
S\"oren Schweers,$^1$, Alexander Antonov,$^1$, Artem Ryabov,$^2$ and Philipp Maass,$^3$\\[2ex]
$^1$\textit{Universit\"{a}t Osnabr\"{u}ck, Fachbereich Physik, Barbarastra{\ss}e 7, D-49076 Osnabr\"uck, Germany}\\[1ex]
$^2$\textit{Charles University, Faculty of Mathematics and Physics, Department of Macromolecular Physics,\\ V Hole\v{s}ovi\v{c}k\'{a}ch 2, 
CZ-18000 Praha 8, Czech Republic}
\end{center}

\setstretch{1.5}
\vspace{1ex}\noindent

In Sec.~\ref{sec:cluster-size-distribution} of this Supplemental Material,
we show that the cluster size distribution of particles with adhesive interactions is a geometric one with
a mean cluster size given by Eq.~(8) in the main text. Section~\ref{sec:msd-scaling-fixed-cluster-sizes} shows the scaling behavior of the mean squared displacement according to Eq.~(11) of the main text, which applies to a system, where the clusters in the initial equilibrium configuration are assumed to have fixed sizes in the course of time. In Sec.~\ref{sec:long-time-subdiffusion} we derive the
coefficient $D_{1/2}$ and discuss its dependence on initial conditions.

\section{Cluster size distribution}
\label{sec:cluster-size-distribution}
We show that for the geometric cluster size distribution (7), the pair correlation function
must have the $\delta$-function singularities for $r=n\sigma$ with the amplitudes 
\begin{equation}
A_n=q^n/\rho 
\label{eq:An}
\end{equation}
as in Eq.~(4). 

Let us first note that $\delta$-function singularities in $g(r)$ can appear if there is a finite probability of finding a particle exactly at 
a distance $r=n\sigma$, $n\in\mathbb{N}$, from another particle. This is possible only if the two particles 
at distance $r$ are in the same cluster.

Taking a reference particle at position zero, $\rho g(x)\dd x$ is the mean number of particles in an infinitesimal interval $\dd x$ at position $x>0$
from the reference particle.
As there can be at most one particle in the infinitesimal interval $\dd x$, 
$\rho g(x)\dd x$ is equal to the probability of finding another particle in an infinitesimal interval $\dd x$ at 
position $x$.  Therefore, $\rho A_n$ is the probability of finding a particle
at $x=n\sigma$ in a cluster containing the reference particle.
Such a particle is present in all cases, 
where a cluster of size $n'\ge n$ is in contact right to the reference particle.
The probability of finding
a sequence of exactly $n'$ particles right to the reference particle is $(1-q)q^{n'}$, $n'=0,1,2\ldots$ Accordingly, we obtain
\begin{equation}
\rho A_n=\sum_{n'=n}^\infty (1-q)q^{n'}=q^n\,,
\end{equation}
in agreement with the $\delta$-function amplitudes in Eq.~\eqref{eq:An}.

By our MC simulations we have verified the geometric distribution (7), see 
Fig.~\ref{fig:cluster_size_distribution}(a) for the case $\rho\sigma=0.25$ and $\gamma=10$.
Analogous results were obtained for other values of $\rho$ and $\gamma$. 
Equation~(8) for the mean cluster size $\bar n$ is also validated by the MC results, see
Fig.~\ref{fig:cluster_size_distribution}(b).

\begin{figure}[t!]
\includegraphics[width=\textwidth]{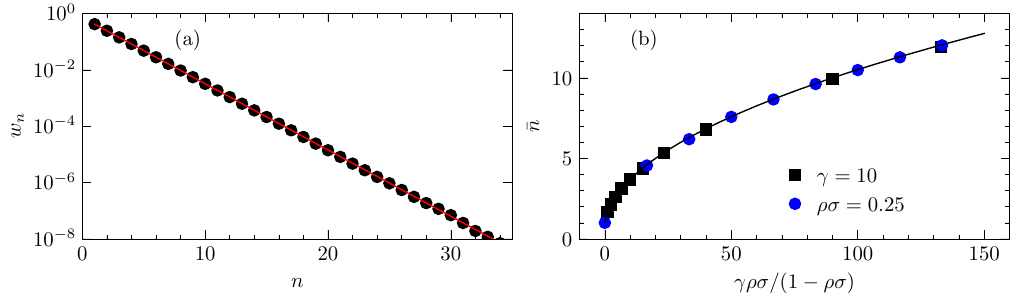}
\caption{(a) Representative example of the cluster size distribution for stickiness $\gamma=10$ and coverage
$\rho\sigma=0.25$. In (b) 
 the mean cluster size is plotted as a function of the effective stickiness $\tilde\gamma=\gamma\rho\sigma/(1-\rho\sigma)$
 for fixed $\gamma=10$ (black squares) and for
 fixed $\rho\sigma=0.25$ (blue circles).
Symbols mark the MC results. The line in (a) displays 
the geometric distribution~(7) with the mean cluster size $\bar n\cong2.39$ from Eq.~(8), and the line in (b) shows
the analytical result in Eq.~(8).}
\label{fig:cluster_size_distribution}
\label{fig:cluster_size_distribution}
\end{figure}

\section{Scaling of mean squared displacement\\ for systems with fixed cluster sizes}
\label{sec:msd-scaling-fixed-cluster-sizes}
\begin{figure}[h!]
\includegraphics[width=\textwidth]{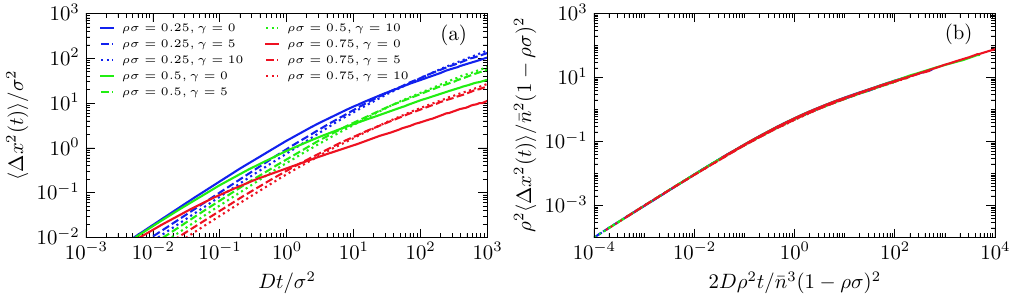}
\caption{(a) Mean squared displacement of a tagged particle for various particle densities $\rho$ and adhesive strengths
$\gamma$, when the cluster sizes are fixed in the initial configuration of the equilibrium ensemble. 
(b) Scaled mean-square displacement vs.\ scaled time according to Eq.~(11) in the main text.}
\label{fig:scaling_MSD_for_fixed_cluster_sizes}
\end{figure}

Figure~\ref{fig:scaling_MSD_for_fixed_cluster_sizes}(a) shows the
mean squared displacement for various particle number densities $\rho$ and adhesive strengths $\gamma$, when the clusters in the initial configuration of the equilibrium ensemble are assumed to have a fixed size, i.e.\ when we neglect the fragmentation and merging of the clusters formed due to the adhesive interaction. When scaling the data according to Eq.~(11) in the main text, 
all data collapse onto a single master curve, see Fig.~\ref{fig:scaling_MSD_for_fixed_cluster_sizes}(b).

\section{Coefficient $D_{1/2}$ in long-time subdiffusion\\ of adhesive particles}
\label{sec:long-time-subdiffusion}
Based on a cumulant expansion of the probability density function for a tagged particle, it has been shown \cite{Kollmann:2003} 
that the mean squared displacement of a tagged particle in a single file system of interacting particles 
has the long-time behavior according to Eq.~(1) in the main text with
\begin{equation}
D_{1/2} = \frac{S(0)}{\rho} \sqrt{\frac{D_{\rm e}}{\pi}}\,.
\label{eq:D1/2-kollmann-1}
\end{equation}
Here $S(0)$ is the static structure factor and $D_{\rm e}$ is a short-time effective diffusion coefficient.
This $D_{\rm e}$  is a collective diffusion coefficient, characterizing the short-time decay of long-wavelength density fluctuations, 
or, equivalently, the diffusive spreading 
of the center of mass at short times.  In the absence of hydrodynamic interactions, it holds \cite{Naegele:1996}
\begin{equation}
D_{\rm e}=\frac{D}{S(0)}\,.
\label{eq:De}
\end{equation}
The  static structure factor is related to the isothermal compressibility $\chi$,
\begin{equation}
S(0)=k_{\rm B}T\rho\,\chi\,.
\label{eq:S(0)-chi}
\end{equation}
Combining Eqs.~\eqref{eq:D1/2-kollmann-1}-\eqref{eq:S(0)-chi} gives
\begin{equation}
D_{1/2}=\sqrt{\frac{k_{\rm B} T D\chi}{\pi \rho}}\,.
\label{eq:D1/2-kollmann-2}
\end{equation}
For the sticky hard sphere fluid in one dimension, 
the isothermal compressibility
is \cite{Yuste/Santos:1993}
\begin{equation}
k_{\rm B}T\rho\,\chi=1+\rho\int\limits_0^\infty\dd r\, [g(r)-1]=(1-\rho\sigma)\sqrt{(1-\rho\sigma)(1-\rho\sigma+4\gamma\rho\sigma)}
=(1-\rho\sigma)^2[2\bar n(\tilde\gamma)-1]\,,
\label{eq:chi-sticky-hard-rods}
\end{equation}
with $\bar n(\tilde\gamma)$ from Eq.~(8) of the main text.
Equations~\eqref{eq:D1/2-kollmann-2} and \eqref{eq:chi-sticky-hard-rods} yield Eq.~(12) in the main text. 

\begin{figure}[t!]
\includegraphics[width=0.6\textwidth]{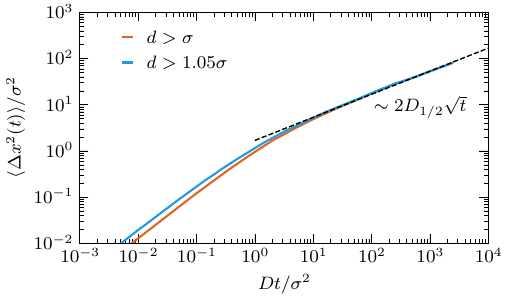}
\caption{Time-dependent mean squared displacement of a tagged particle in an equilibrated one-dimensional system of 
sticky hard spheres at stickiness $\gamma=1$ and coverage $\rho\sigma=0.5$. The simulation results are shown for two 
initial conditions: in the first case (red line), the tagged particle is a single particle (not part of a cluster), implying that its 
distance $d$ to any other particle is larger than $\sigma$. In the second case (blue line), $d>1.05\,\sigma$.  At long times,
the mean squared displacement behaves as $\langle \Delta x^2(t)\rangle\sim 2D_{1/2}\sqrt{t}$, where $D_{1/2}$ 
is independent of the initial condition and has a value $D_{1/2}\cong0.844$, in agreement with Eq.~(12) of the main text.
The asymptotic behavior is indicated by the dashed line.}
\label{fig:MSD_for_other_inital_conditions_of_tagged_particle}
\end{figure}

Equation~\eqref{eq:D1/2-kollmann-1} implies that $D_{1/2}$ depends only on the structure factor and particle number density, i.e.\ it is 
independent of the initial position of the tagged particle for given $S(0)$ and $\rho$. We exemplify this independence on the initial 
conditions for two cases in Fig.~\ref{fig:MSD_for_other_inital_conditions_of_tagged_particle}: in the first case, the tagged particle is a single 
particle, i.e.\ not part of a cluster (red line), and in the second case, the single particle has a distance larger than $1.05\sigma$ 
from any other particle. In both cases, the overall structure is that of an equilibrated system of adhesive particles with stickiness
$\gamma=1$ and coverage $\rho\sigma=0.5$. As can be seen from the figure, $D_{1/2}$ is the same for both initial conditions (and the 
same for an equilibrated tagged particle). By contrast, the short-time diffusion coefficient depends on the initial distribution of starting 
positions of the tagged particle.


\begin{thebibliography}{54}%
\makeatletter
\providecommand \@ifxundefined [1]{%
 \@ifx{#1\undefined}
}%
\providecommand \@ifnum [1]{%
 \ifnum #1\expandafter \@firstoftwo
 \else \expandafter \@secondoftwo
 \fi
}%
\providecommand \@ifx [1]{%
 \ifx #1\expandafter \@firstoftwo
 \else \expandafter \@secondoftwo
 \fi
}%
\providecommand \natexlab [1]{#1}%
\providecommand \enquote  [1]{``#1''}%
\providecommand \bibnamefont  [1]{#1}%
\providecommand \bibfnamefont [1]{#1}%
\providecommand \citenamefont [1]{#1}%
\providecommand \href@noop [0]{\@secondoftwo}%
\providecommand \href [0]{\begingroup \@sanitize@url \@href}%
\providecommand \@href[1]{\@@startlink{#1}\@@href}%
\providecommand \@@href[1]{\endgroup#1\@@endlink}%
\providecommand \@sanitize@url [0]{\catcode `\\12\catcode `\$12\catcode
  `\&12\catcode `\#12\catcode `\^12\catcode `\_12\catcode `\%12\relax}%
\providecommand \@@startlink[1]{}%
\providecommand \@@endlink[0]{}%
\providecommand \url  [0]{\begingroup\@sanitize@url \@url }%
\providecommand \@url [1]{\endgroup\@href {#1}{\urlprefix }}%
\providecommand \urlprefix  [0]{URL }%
\providecommand \Eprint [0]{\href }%
\providecommand \doibase [0]{https://doi.org/}%
\providecommand \selectlanguage [0]{\@gobble}%
\providecommand \bibinfo  [0]{\@secondoftwo}%
\providecommand \bibfield  [0]{\@secondoftwo}%
\providecommand \translation [1]{[#1]}%
\providecommand \BibitemOpen [0]{}%
\providecommand \bibitemStop [0]{}%
\providecommand \bibitemNoStop [0]{.\EOS\space}%
\providecommand \EOS [0]{\spacefactor3000\relax}%
\providecommand \BibitemShut  [1]{\csname bibitem#1\endcsname}%
\let\auto@bib@innerbib\@empty
\bibitem [{\citenamefont {K\"arger}(2015)}]{Kaerger:2014}%
  \BibitemOpen
  \bibfield  {author} {\bibinfo {author} {\bibfnamefont {J.}~\bibnamefont
  {K\"arger}},\ }\bibfield  {title} {\bibinfo {title} {Transport phenomena in
  nanoporous materials},\ }\href
  {https://doi.org/https://doi.org/10.1002/cphc.201402340} {\bibfield
  {journal} {\bibinfo  {journal} {ChemPhysChem}\ }\textbf {\bibinfo {volume}
  {16}},\ \bibinfo {pages} {24} (\bibinfo {year} {2015})}\BibitemShut {NoStop}%
\bibitem [{\citenamefont {Nyg\r{a}rd}(2017)}]{Nygard:2017}%
  \BibitemOpen
  \bibfield  {author} {\bibinfo {author} {\bibfnamefont {K.}~\bibnamefont
  {Nyg\r{a}rd}},\ }\bibfield  {title} {\bibinfo {title} {Colloidal diffusion in
  confined geometries},\ }\href {https://doi.org/10.1039/C7CP02497E} {\bibfield
   {journal} {\bibinfo  {journal} {Phys. Chem. Chem. Phys.}\ }\textbf {\bibinfo
  {volume} {19}},\ \bibinfo {pages} {23632} (\bibinfo {year}
  {2017})}\BibitemShut {NoStop}%
\bibitem [{\citenamefont {Taloni}\ \emph {et~al.}(2017)\citenamefont {Taloni},
  \citenamefont {Flomenbom}, \citenamefont {Casta{\~n}eda-Priego},\ and\
  \citenamefont {Marchesoni}}]{Taloni/etal:2017}%
  \BibitemOpen
  \bibfield  {author} {\bibinfo {author} {\bibfnamefont {A.}~\bibnamefont
  {Taloni}}, \bibinfo {author} {\bibfnamefont {O.}~\bibnamefont {Flomenbom}},
  \bibinfo {author} {\bibfnamefont {R.}~\bibnamefont {Casta{\~n}eda-Priego}},\
  and\ \bibinfo {author} {\bibfnamefont {F.}~\bibnamefont {Marchesoni}},\
  }\bibfield  {title} {\bibinfo {title} {Single file dynamics in soft
  materials},\ }\href {https://doi.org/10.1039/C6SM02570F} {\bibfield
  {journal} {\bibinfo  {journal} {Soft Matter}\ }\textbf {\bibinfo {volume}
  {13}},\ \bibinfo {pages} {1096} (\bibinfo {year} {2017})}\BibitemShut
  {NoStop}%
\bibitem [{\citenamefont {Bukowski}\ \emph {et~al.}(2021)\citenamefont
  {Bukowski}, \citenamefont {Keil}, \citenamefont {Ravikovitch}, \citenamefont
  {Sastre}, \citenamefont {Snurr},\ and\ \citenamefont
  {Coppens}}]{Bukowski/etal:2021}%
  \BibitemOpen
  \bibfield  {author} {\bibinfo {author} {\bibfnamefont {B.~C.}\ \bibnamefont
  {Bukowski}}, \bibinfo {author} {\bibfnamefont {F.~J.}\ \bibnamefont {Keil}},
  \bibinfo {author} {\bibfnamefont {P.~I.}\ \bibnamefont {Ravikovitch}},
  \bibinfo {author} {\bibfnamefont {G.}~\bibnamefont {Sastre}}, \bibinfo
  {author} {\bibfnamefont {R.~Q.}\ \bibnamefont {Snurr}},\ and\ \bibinfo
  {author} {\bibfnamefont {M.-O.}\ \bibnamefont {Coppens}},\ }\bibfield
  {title} {\bibinfo {title} {Connecting theory and simulation with experiment
  for the study of diffusion in nanoporous solids},\ }\href
  {https://doi.org/10.1007/s10450-021-00314-y} {\bibfield  {journal} {\bibinfo
  {journal} {Adsorption}\ }\textbf {\bibinfo {volume} {27}},\ \bibinfo {pages}
  {683} (\bibinfo {year} {2021})}\BibitemShut {NoStop}%
\bibitem [{\citenamefont {Hahn}\ \emph {et~al.}(1996)\citenamefont {Hahn},
  \citenamefont {K\"arger},\ and\ \citenamefont {Kukla}}]{Hahn/etal:1996}%
  \BibitemOpen
  \bibfield  {author} {\bibinfo {author} {\bibfnamefont {K.}~\bibnamefont
  {Hahn}}, \bibinfo {author} {\bibfnamefont {J.}~\bibnamefont {K\"arger}},\
  and\ \bibinfo {author} {\bibfnamefont {V.}~\bibnamefont {Kukla}},\ }\bibfield
   {title} {\bibinfo {title} {Single-file diffusion observation},\ }\href
  {https://doi.org/10.1103/PhysRevLett.76.2762} {\bibfield  {journal} {\bibinfo
   {journal} {Phys. Rev. Lett.}\ }\textbf {\bibinfo {volume} {76}},\ \bibinfo
  {pages} {2762} (\bibinfo {year} {1996})}\BibitemShut {NoStop}%
\bibitem [{\citenamefont {Hahn}\ and\ \citenamefont
  {K\"arger}(1998)}]{Hahn/Kaerger:1998}%
  \BibitemOpen
  \bibfield  {author} {\bibinfo {author} {\bibfnamefont {K.}~\bibnamefont
  {Hahn}}\ and\ \bibinfo {author} {\bibfnamefont {J.}~\bibnamefont
  {K\"arger}},\ }\bibfield  {title} {\bibinfo {title} {Deviations from the
  normal time regime of single-file diffusion},\ }\href
  {https://doi.org/10.1021/jp981039h} {\bibfield  {journal} {\bibinfo
  {journal} {J. Phys. Chem. B}\ }\textbf {\bibinfo {volume} {102}},\ \bibinfo
  {pages} {5766} (\bibinfo {year} {1998})}\BibitemShut {NoStop}%
\bibitem [{\citenamefont {Chmelik}\ \emph {et~al.}(2018)\citenamefont
  {Chmelik}, \citenamefont {Caro}, \citenamefont {Freude}, \citenamefont
  {Haase}, \citenamefont {Valiullin},\ and\ \citenamefont
  {K\"arger}}]{Chmelik/etal:2018}%
  \BibitemOpen
  \bibfield  {author} {\bibinfo {author} {\bibfnamefont {C.}~\bibnamefont
  {Chmelik}}, \bibinfo {author} {\bibfnamefont {J.}~\bibnamefont {Caro}},
  \bibinfo {author} {\bibfnamefont {D.}~\bibnamefont {Freude}}, \bibinfo
  {author} {\bibfnamefont {J.}~\bibnamefont {Haase}}, \bibinfo {author}
  {\bibfnamefont {R.}~\bibnamefont {Valiullin}},\ and\ \bibinfo {author}
  {\bibfnamefont {J.}~\bibnamefont {K\"arger}},\ }\bibfield  {title} {\bibinfo
  {title} {Diffusive {S}preading of {M}olecules in {N}anoporous {M}aterials},\
  }in\ \href {https://doi.org/10.1007/978-3-319-67798-9} {\emph {\bibinfo
  {booktitle} {Diffusive {S}preading in {N}ature, {T}echnology and
  {S}ociety}}},\ \bibinfo {editor} {edited by\ \bibinfo {editor} {\bibfnamefont
  {A.}~\bibnamefont {Bunde}}, \bibinfo {editor} {\bibfnamefont
  {J.}~\bibnamefont {Caro}}, \bibinfo {editor} {\bibfnamefont {J.}~\bibnamefont
  {K\"arger}},\ and\ \bibinfo {editor} {\bibfnamefont {G.}~\bibnamefont
  {Vogl}}}\ (\bibinfo  {publisher} {Springer International Publishing},\
  \bibinfo {address} {Cham},\ \bibinfo {year} {2018})\ Chap.~\bibinfo {chapter}
  {10}, pp.\ \bibinfo {pages} {171--202}\BibitemShut {NoStop}%
\bibitem [{\citenamefont {Wei}\ \emph {et~al.}(2000)\citenamefont {Wei},
  \citenamefont {Bechinger},\ and\ \citenamefont {Leiderer}}]{Wei/etal:2000}%
  \BibitemOpen
  \bibfield  {author} {\bibinfo {author} {\bibfnamefont {Q.-H.}\ \bibnamefont
  {Wei}}, \bibinfo {author} {\bibfnamefont {C.}~\bibnamefont {Bechinger}},\
  and\ \bibinfo {author} {\bibfnamefont {P.}~\bibnamefont {Leiderer}},\
  }\bibfield  {title} {\bibinfo {title} {Single-file diffusion of colloids in
  one-dimensional channels},\ }\href
  {https://doi.org/10.1126/science.287.5453.625} {\bibfield  {journal}
  {\bibinfo  {journal} {Science}\ }\textbf {\bibinfo {volume} {287}},\ \bibinfo
  {pages} {625} (\bibinfo {year} {2000})}\BibitemShut {NoStop}%
\bibitem [{\citenamefont {Cui}\ \emph {et~al.}(2002)\citenamefont {Cui},
  \citenamefont {Diamant},\ and\ \citenamefont {Lin}}]{Cui/etal:2002}%
  \BibitemOpen
  \bibfield  {author} {\bibinfo {author} {\bibfnamefont {B.}~\bibnamefont
  {Cui}}, \bibinfo {author} {\bibfnamefont {H.}~\bibnamefont {Diamant}},\ and\
  \bibinfo {author} {\bibfnamefont {B.}~\bibnamefont {Lin}},\ }\bibfield
  {title} {\bibinfo {title} {Screened hydrodynamic interaction in a narrow
  channel},\ }\href {https://doi.org/10.1103/PhysRevLett.89.188302} {\bibfield
  {journal} {\bibinfo  {journal} {Phys. Rev. Lett.}\ }\textbf {\bibinfo
  {volume} {89}},\ \bibinfo {pages} {188302} (\bibinfo {year}
  {2002})}\BibitemShut {NoStop}%
\bibitem [{\citenamefont {Lin}\ \emph {et~al.}(2002)\citenamefont {Lin},
  \citenamefont {Cui}, \citenamefont {Lee},\ and\ \citenamefont
  {Yu}}]{Lin/etal:2002}%
  \BibitemOpen
  \bibfield  {author} {\bibinfo {author} {\bibfnamefont {B.}~\bibnamefont
  {Lin}}, \bibinfo {author} {\bibfnamefont {B.}~\bibnamefont {Cui}}, \bibinfo
  {author} {\bibfnamefont {J.-H.}\ \bibnamefont {Lee}},\ and\ \bibinfo {author}
  {\bibfnamefont {J.}~\bibnamefont {Yu}},\ }\bibfield  {title} {\bibinfo
  {title} {Hydrodynamic coupling in diffusion of quasi--one-dimensional
  brownian particles},\ }\href {https://doi.org/10.1209/epl/i2002-00523-2}
  {\bibfield  {journal} {\bibinfo  {journal} {Europhys. Lett. ({EPL})}\
  }\textbf {\bibinfo {volume} {57}},\ \bibinfo {pages} {724} (\bibinfo {year}
  {2002})}\BibitemShut {NoStop}%
\bibitem [{\citenamefont {Lutz}\ \emph
  {et~al.}(2004{\natexlab{a}})\citenamefont {Lutz}, \citenamefont {Kollmann},\
  and\ \citenamefont {Bechinger}}]{Lutz/etal:2004a}%
  \BibitemOpen
  \bibfield  {author} {\bibinfo {author} {\bibfnamefont {C.}~\bibnamefont
  {Lutz}}, \bibinfo {author} {\bibfnamefont {M.}~\bibnamefont {Kollmann}},\
  and\ \bibinfo {author} {\bibfnamefont {C.}~\bibnamefont {Bechinger}},\
  }\bibfield  {title} {\bibinfo {title} {Single-{{File Diffusion}} of
  {{Colloids}} in {{One}}-{{Dimensional Channels}}},\ }\href
  {https://doi.org/10.1103/PhysRevLett.93.026001} {\bibfield  {journal}
  {\bibinfo  {journal} {Phys. Rev. Lett.}\ }\textbf {\bibinfo {volume} {93}},\
  \bibinfo {pages} {026001} (\bibinfo {year} {2004}{\natexlab{a}})}\BibitemShut
  {NoStop}%
\bibitem [{\citenamefont {Lutz}\ \emph
  {et~al.}(2004{\natexlab{b}})\citenamefont {Lutz}, \citenamefont {Kollmann},
  \citenamefont {Leiderer},\ and\ \citenamefont {Bechinger}}]{Lutz/etal:2004b}%
  \BibitemOpen
  \bibfield  {author} {\bibinfo {author} {\bibfnamefont {C.}~\bibnamefont
  {Lutz}}, \bibinfo {author} {\bibfnamefont {M.}~\bibnamefont {Kollmann}},
  \bibinfo {author} {\bibfnamefont {P.}~\bibnamefont {Leiderer}},\ and\
  \bibinfo {author} {\bibfnamefont {C.}~\bibnamefont {Bechinger}},\ }\bibfield
  {title} {\bibinfo {title} {Diffusion of colloids in one-dimensional light
  channels},\ }\href {https://doi.org/10.1088/0953-8984/16/38/022} {\bibfield
  {journal} {\bibinfo  {journal} {J. Phys.: Condens. Mat.}\ }\textbf {\bibinfo
  {volume} {16}},\ \bibinfo {pages} {S4075} (\bibinfo {year}
  {2004}{\natexlab{b}})}\BibitemShut {NoStop}%
\bibitem [{\citenamefont {Lin}\ \emph {et~al.}(2005)\citenamefont {Lin},
  \citenamefont {Meron}, \citenamefont {Cui}, \citenamefont {Rice},\ and\
  \citenamefont {Diamant}}]{Lin/etal:2005}%
  \BibitemOpen
  \bibfield  {author} {\bibinfo {author} {\bibfnamefont {B.}~\bibnamefont
  {Lin}}, \bibinfo {author} {\bibfnamefont {M.}~\bibnamefont {Meron}}, \bibinfo
  {author} {\bibfnamefont {B.}~\bibnamefont {Cui}}, \bibinfo {author}
  {\bibfnamefont {S.~A.}\ \bibnamefont {Rice}},\ and\ \bibinfo {author}
  {\bibfnamefont {H.}~\bibnamefont {Diamant}},\ }\bibfield  {title} {\bibinfo
  {title} {From random walk to single-file diffusion},\ }\href
  {https://doi.org/10.1103/PhysRevLett.94.216001} {\bibfield  {journal}
  {\bibinfo  {journal} {Phys. Rev. Lett.}\ }\textbf {\bibinfo {volume} {94}},\
  \bibinfo {pages} {216001} (\bibinfo {year} {2005})}\BibitemShut {NoStop}%
\bibitem [{\citenamefont {K\"oppl}\ \emph {et~al.}(2006)\citenamefont
  {K\"oppl}, \citenamefont {Henseler}, \citenamefont {Erbe}, \citenamefont
  {Nielaba},\ and\ \citenamefont {Leiderer}}]{Koeppl/etal:2006}%
  \BibitemOpen
  \bibfield  {author} {\bibinfo {author} {\bibfnamefont {M.}~\bibnamefont
  {K\"oppl}}, \bibinfo {author} {\bibfnamefont {P.}~\bibnamefont {Henseler}},
  \bibinfo {author} {\bibfnamefont {A.}~\bibnamefont {Erbe}}, \bibinfo {author}
  {\bibfnamefont {P.}~\bibnamefont {Nielaba}},\ and\ \bibinfo {author}
  {\bibfnamefont {P.}~\bibnamefont {Leiderer}},\ }\bibfield  {title} {\bibinfo
  {title} {Layer reduction in driven 2d-colloidal systems through
  microchannels},\ }\href {https://doi.org/10.1103/PhysRevLett.97.208302}
  {\bibfield  {journal} {\bibinfo  {journal} {Phys. Rev. Lett.}\ }\textbf
  {\bibinfo {volume} {97}},\ \bibinfo {pages} {208302} (\bibinfo {year}
  {2006})}\BibitemShut {NoStop}%
\bibitem [{\citenamefont {Henseler}\ \emph {et~al.}(2010)\citenamefont
  {Henseler}, \citenamefont {Erbe}, \citenamefont {K\"oppl}, \citenamefont
  {Leiderer},\ and\ \citenamefont {Nielaba}}]{Henseler/etal:2010}%
  \BibitemOpen
  \bibfield  {author} {\bibinfo {author} {\bibfnamefont {P.}~\bibnamefont
  {Henseler}}, \bibinfo {author} {\bibfnamefont {A.}~\bibnamefont {Erbe}},
  \bibinfo {author} {\bibfnamefont {M.}~\bibnamefont {K\"oppl}}, \bibinfo
  {author} {\bibfnamefont {P.}~\bibnamefont {Leiderer}},\ and\ \bibinfo
  {author} {\bibfnamefont {P.}~\bibnamefont {Nielaba}},\ }\bibfield  {title}
  {\bibinfo {title} {Density reduction and diffusion in driven two-dimensional
  colloidal systems through microchannels},\ }\href
  {https://doi.org/10.1103/PhysRevE.81.041402} {\bibfield  {journal} {\bibinfo
  {journal} {Phys. Rev. E}\ }\textbf {\bibinfo {volume} {81}},\ \bibinfo
  {pages} {041402} (\bibinfo {year} {2010})}\BibitemShut {NoStop}%
\bibitem [{\citenamefont {Cheng}\ and\ \citenamefont
  {Bowers}(2007)}]{Cheng/Bowers:2007}%
  \BibitemOpen
  \bibfield  {author} {\bibinfo {author} {\bibfnamefont {C.-Y.}\ \bibnamefont
  {Cheng}}\ and\ \bibinfo {author} {\bibfnamefont {C.~R.}\ \bibnamefont
  {Bowers}},\ }\bibfield  {title} {\bibinfo {title} {Observation of single-file
  diffusion in dipeptide nanotubes by continuous-flow hyperpolarized
  {X}enon-129 {NMR} spectroscopy},\ }\href
  {https://doi.org/10.1002/cphc.200700336} {\bibfield  {journal} {\bibinfo
  {journal} {ChemPhysChem}\ }\textbf {\bibinfo {volume} {8}},\ \bibinfo {pages}
  {2077} (\bibinfo {year} {2007})}\BibitemShut {NoStop}%
\bibitem [{\citenamefont {Das}\ \emph {et~al.}(2010)\citenamefont {Das},
  \citenamefont {Jayanthi}, \citenamefont {Deepak}, \citenamefont {Ramanathan},
  \citenamefont {Kumar}, \citenamefont {Dasgupta},\ and\ \citenamefont
  {Sood}}]{Das/etal:2010}%
  \BibitemOpen
  \bibfield  {author} {\bibinfo {author} {\bibfnamefont {A.}~\bibnamefont
  {Das}}, \bibinfo {author} {\bibfnamefont {S.}~\bibnamefont {Jayanthi}},
  \bibinfo {author} {\bibfnamefont {H.~S. M.~V.}\ \bibnamefont {Deepak}},
  \bibinfo {author} {\bibfnamefont {K.~V.}\ \bibnamefont {Ramanathan}},
  \bibinfo {author} {\bibfnamefont {A.}~\bibnamefont {Kumar}}, \bibinfo
  {author} {\bibfnamefont {C.}~\bibnamefont {Dasgupta}},\ and\ \bibinfo
  {author} {\bibfnamefont {A.~K.}\ \bibnamefont {Sood}},\ }\bibfield  {title}
  {\bibinfo {title} {Single-file diffusion of confined water inside {SWNT}s: An
  {NMR} study},\ }\href {https://doi.org/10.1021/nn901554h} {\bibfield
  {journal} {\bibinfo  {journal} {ACS Nano}\ }\textbf {\bibinfo {volume} {4}},\
  \bibinfo {pages} {1687} (\bibinfo {year} {2010})}\BibitemShut {NoStop}%
\bibitem [{\citenamefont {Dvoyashkin}\ \emph {et~al.}(2014)\citenamefont
  {Dvoyashkin}, \citenamefont {Bhase}, \citenamefont {Mirnazari}, \citenamefont
  {Vasenkov},\ and\ \citenamefont {Bowers}}]{Dvoyashkin/etal:2014}%
  \BibitemOpen
  \bibfield  {author} {\bibinfo {author} {\bibfnamefont {M.}~\bibnamefont
  {Dvoyashkin}}, \bibinfo {author} {\bibfnamefont {H.}~\bibnamefont {Bhase}},
  \bibinfo {author} {\bibfnamefont {N.}~\bibnamefont {Mirnazari}}, \bibinfo
  {author} {\bibfnamefont {S.}~\bibnamefont {Vasenkov}},\ and\ \bibinfo
  {author} {\bibfnamefont {C.~R.}\ \bibnamefont {Bowers}},\ }\bibfield  {title}
  {\bibinfo {title} {Single-file nanochannel persistence lengths from {NMR}},\
  }\href {https://doi.org/10.1021/ac403868t} {\bibfield  {journal} {\bibinfo
  {journal} {Anal. Chem.}\ }\textbf {\bibinfo {volume} {86}},\ \bibinfo {pages}
  {2200} (\bibinfo {year} {2014})}\BibitemShut {NoStop}%
\bibitem [{\citenamefont {Cao}\ \emph {et~al.}(2018)\citenamefont {Cao},
  \citenamefont {Huang}, \citenamefont {Ma}, \citenamefont {Lu},\ and\
  \citenamefont {Lu}}]{Cao/etal:2018}%
  \BibitemOpen
  \bibfield  {author} {\bibinfo {author} {\bibfnamefont {W.}~\bibnamefont
  {Cao}}, \bibinfo {author} {\bibfnamefont {L.}~\bibnamefont {Huang}}, \bibinfo
  {author} {\bibfnamefont {M.}~\bibnamefont {Ma}}, \bibinfo {author}
  {\bibfnamefont {L.}~\bibnamefont {Lu}},\ and\ \bibinfo {author}
  {\bibfnamefont {X.}~\bibnamefont {Lu}},\ }\bibfield  {title} {\bibinfo
  {title} {Water in narrow carbon nanotubes: Roughness promoted diffusion
  transition},\ }\href {https://doi.org/10.1021/acs.jpcc.8b02929} {\bibfield
  {journal} {\bibinfo  {journal} {J. Phys. Chem. C}\ }\textbf {\bibinfo
  {volume} {122}},\ \bibinfo {pages} {19124} (\bibinfo {year}
  {2018})}\BibitemShut {NoStop}%
\bibitem [{\citenamefont {Zeng}\ \emph {et~al.}(2018)\citenamefont {Zeng},
  \citenamefont {Chen}, \citenamefont {Wang}, \citenamefont {Zhou},
  \citenamefont {Chen},\ and\ \citenamefont {Dai}}]{Zeng/etal:2018}%
  \BibitemOpen
  \bibfield  {author} {\bibinfo {author} {\bibfnamefont {S.}~\bibnamefont
  {Zeng}}, \bibinfo {author} {\bibfnamefont {J.}~\bibnamefont {Chen}}, \bibinfo
  {author} {\bibfnamefont {X.}~\bibnamefont {Wang}}, \bibinfo {author}
  {\bibfnamefont {G.}~\bibnamefont {Zhou}}, \bibinfo {author} {\bibfnamefont
  {L.}~\bibnamefont {Chen}},\ and\ \bibinfo {author} {\bibfnamefont
  {C.}~\bibnamefont {Dai}},\ }\bibfield  {title} {\bibinfo {title} {Selective
  transport through the ultrashort carbon nanotubes embedded in lipid
  bilayers},\ }\href {https://doi.org/10.1021/acs.jpcc.8b07861} {\bibfield
  {journal} {\bibinfo  {journal} {J. Phys. Chem. C}\ }\textbf {\bibinfo
  {volume} {122}},\ \bibinfo {pages} {27681} (\bibinfo {year}
  {2018})}\BibitemShut {NoStop}%
\bibitem [{\citenamefont {Bauer}\ and\ \citenamefont
  {Nadler}(2006)}]{Bauer/Nadler:2006}%
  \BibitemOpen
  \bibfield  {author} {\bibinfo {author} {\bibfnamefont {W.~R.}\ \bibnamefont
  {Bauer}}\ and\ \bibinfo {author} {\bibfnamefont {W.}~\bibnamefont {Nadler}},\
  }\bibfield  {title} {\bibinfo {title} {Molecular transport through channels
  and pores: Effects of in-channel interactions and blocking},\ }\href
  {https://doi.org/10.1073/pnas.0601769103} {\bibfield  {journal} {\bibinfo
  {journal} {Proc. Natl. Acad. Sci. U.S.A.}\ }\textbf {\bibinfo {volume}
  {103}},\ \bibinfo {pages} {11446} (\bibinfo {year} {2006})}\BibitemShut
  {NoStop}%
\bibitem [{\citenamefont {Kahms}\ \emph {et~al.}(2009)\citenamefont {Kahms},
  \citenamefont {Lehrich}, \citenamefont {H{\"u}ve}, \citenamefont {Sanetra},\
  and\ \citenamefont {Peters}}]{Kahms/etal:2009}%
  \BibitemOpen
  \bibfield  {author} {\bibinfo {author} {\bibfnamefont {M.}~\bibnamefont
  {Kahms}}, \bibinfo {author} {\bibfnamefont {P.}~\bibnamefont {Lehrich}},
  \bibinfo {author} {\bibfnamefont {J.}~\bibnamefont {H{\"u}ve}}, \bibinfo
  {author} {\bibfnamefont {N.}~\bibnamefont {Sanetra}},\ and\ \bibinfo {author}
  {\bibfnamefont {R.}~\bibnamefont {Peters}},\ }\bibfield  {title} {\bibinfo
  {title} {Binding site distribution of nuclear transport receptors and
  transport complexes in single nuclear pore complexes},\ }\href
  {https://doi.org/https://doi.org/10.1111/j.1600-0854.2009.00947.x} {\bibfield
   {journal} {\bibinfo  {journal} {Traffic}\ }\textbf {\bibinfo {volume}
  {10}},\ \bibinfo {pages} {1228} (\bibinfo {year} {2009})}\BibitemShut
  {NoStop}%
\bibitem [{\citenamefont {Yang}\ \emph {et~al.}(2010)\citenamefont {Yang},
  \citenamefont {Yang}, \citenamefont {Kim}, \citenamefont {Jeon},
  \citenamefont {Oh}, \citenamefont {Choi}, \citenamefont {Hahn},\ and\
  \citenamefont {Kim}}]{Yang/etal:2010}%
  \BibitemOpen
  \bibfield  {author} {\bibinfo {author} {\bibfnamefont {S.~Y.}\ \bibnamefont
  {Yang}}, \bibinfo {author} {\bibfnamefont {J.-A.}\ \bibnamefont {Yang}},
  \bibinfo {author} {\bibfnamefont {E.-S.}\ \bibnamefont {Kim}}, \bibinfo
  {author} {\bibfnamefont {G.}~\bibnamefont {Jeon}}, \bibinfo {author}
  {\bibfnamefont {E.~J.}\ \bibnamefont {Oh}}, \bibinfo {author} {\bibfnamefont
  {K.~Y.}\ \bibnamefont {Choi}}, \bibinfo {author} {\bibfnamefont {S.~K.}\
  \bibnamefont {Hahn}},\ and\ \bibinfo {author} {\bibfnamefont {J.~K.}\
  \bibnamefont {Kim}},\ }\bibfield  {title} {\bibinfo {title} {Single-file
  diffusion of protein drugs through cylindrical nanochannels},\ }\href
  {https://doi.org/10.1021/nn100464u} {\bibfield  {journal} {\bibinfo
  {journal} {ACS Nano}\ }\textbf {\bibinfo {volume} {4}},\ \bibinfo {pages}
  {3817} (\bibinfo {year} {2010})}\BibitemShut {NoStop}%
\bibitem [{\citenamefont {Luan}\ and\ \citenamefont
  {Zhou}(2018)}]{Luan/Zhou:2018}%
  \BibitemOpen
  \bibfield  {author} {\bibinfo {author} {\bibfnamefont {B.}~\bibnamefont
  {Luan}}\ and\ \bibinfo {author} {\bibfnamefont {R.}~\bibnamefont {Zhou}},\
  }\bibfield  {title} {\bibinfo {title} {Single-file protein translocations
  through graphene--mos2 heterostructure nanopores},\ }\href
  {https://doi.org/10.1021/acs.jpclett.8b01340} {\bibfield  {journal} {\bibinfo
   {journal} {J. Phys. Chem. Lett.}\ }\textbf {\bibinfo {volume} {9}},\
  \bibinfo {pages} {3409} (\bibinfo {year} {2018})}\BibitemShut {NoStop}%
\bibitem [{\citenamefont {Zhao}\ \emph {et~al.}(2018)\citenamefont {Zhao},
  \citenamefont {Wu},\ and\ \citenamefont {Su}}]{Zhao/etal:2018}%
  \BibitemOpen
  \bibfield  {author} {\bibinfo {author} {\bibfnamefont {M.}~\bibnamefont
  {Zhao}}, \bibinfo {author} {\bibfnamefont {W.}~\bibnamefont {Wu}},\ and\
  \bibinfo {author} {\bibfnamefont {B.}~\bibnamefont {Su}},\ }\bibfield
  {title} {\bibinfo {title} {ph-controlled drug release by diffusion through
  silica nanochannel membranes},\ }\href
  {https://doi.org/10.1021/acsami.8b12200} {\bibfield  {journal} {\bibinfo
  {journal} {ACS Appl. Mater. Interfaces}\ }\textbf {\bibinfo {volume} {10}},\
  \bibinfo {pages} {33986} (\bibinfo {year} {2018})}\BibitemShut {NoStop}%
\bibitem [{\citenamefont {K\"arger}\ \emph {et~al.}(2021)\citenamefont
  {K\"arger}, \citenamefont {Ruthven},\ and\ \citenamefont
  {Valiullin}}]{Kaerger/etal:2021}%
  \BibitemOpen
  \bibfield  {author} {\bibinfo {author} {\bibfnamefont {J.}~\bibnamefont
  {K\"arger}}, \bibinfo {author} {\bibfnamefont {D.~M.}\ \bibnamefont
  {Ruthven}},\ and\ \bibinfo {author} {\bibfnamefont {R.}~\bibnamefont
  {Valiullin}},\ }\bibfield  {title} {\bibinfo {title} {Diffusion in nanopores:
  inspecting the grounds},\ }\href {https://doi.org/10.1007/s10450-020-00277-6}
  {\bibfield  {journal} {\bibinfo  {journal} {Adsorption}\ }\textbf {\bibinfo
  {volume} {27}},\ \bibinfo {pages} {267} (\bibinfo {year} {2021})}\BibitemShut
  {NoStop}%
\bibitem [{\citenamefont {Coste}\ \emph {et~al.}(2010)\citenamefont {Coste},
  \citenamefont {Delfau}, \citenamefont {Even},\ and\ \citenamefont
  {Saint~Jean}}]{Coste/etal:2010}%
  \BibitemOpen
  \bibfield  {author} {\bibinfo {author} {\bibfnamefont {C.}~\bibnamefont
  {Coste}}, \bibinfo {author} {\bibfnamefont {J.-B.}\ \bibnamefont {Delfau}},
  \bibinfo {author} {\bibfnamefont {C.}~\bibnamefont {Even}},\ and\ \bibinfo
  {author} {\bibfnamefont {M.}~\bibnamefont {Saint~Jean}},\ }\bibfield  {title}
  {\bibinfo {title} {Single-file diffusion of macroscopic charged particles},\
  }\href {https://doi.org/10.1103/PhysRevE.81.051201} {\bibfield  {journal}
  {\bibinfo  {journal} {Phys. Rev. E}\ }\textbf {\bibinfo {volume} {81}},\
  \bibinfo {pages} {051201} (\bibinfo {year} {2010})}\BibitemShut {NoStop}%
\bibitem [{\citenamefont {Harris}(1965)}]{Harris:1965}%
  \BibitemOpen
  \bibfield  {author} {\bibinfo {author} {\bibfnamefont {T.~E.}\ \bibnamefont
  {Harris}},\ }\bibfield  {title} {\bibinfo {title} {Diffusion with
  ``collisions" between particles},\ }\href {https://doi.org/10.2307/3212197}
  {\bibfield  {journal} {\bibinfo  {journal} {J. Appl. Prob.}\ }\textbf
  {\bibinfo {volume} {2}},\ \bibinfo {pages} {323} (\bibinfo {year}
  {1965})}\BibitemShut {NoStop}%
\bibitem [{\citenamefont {Levitt}(1973)}]{Levitt:1973}%
  \BibitemOpen
  \bibfield  {author} {\bibinfo {author} {\bibfnamefont {D.~G.}\ \bibnamefont
  {Levitt}},\ }\bibfield  {title} {\bibinfo {title} {Dynamics of a single-file
  pore: {N}on-{F}ickian behavior},\ }\href
  {https://doi.org/10.1103/PhysRevA.8.3050} {\bibfield  {journal} {\bibinfo
  {journal} {Phys. Rev. A}\ }\textbf {\bibinfo {volume} {8}},\ \bibinfo {pages}
  {3050} (\bibinfo {year} {1973})}\BibitemShut {NoStop}%
\bibitem [{\citenamefont {Arratia}(1983)}]{Arratia:1983}%
  \BibitemOpen
  \bibfield  {author} {\bibinfo {author} {\bibfnamefont {R.}~\bibnamefont
  {Arratia}},\ }\bibfield  {title} {\bibinfo {title} {The motion of a tagged
  particle in the simple symmetric exclusion system on $\mathbb{Z}^1$},\ }\href
  {https://doi.org/10.1214/aop/1176993602} {\bibfield  {journal} {\bibinfo
  {journal} {Ann. Probab.}\ }\textbf {\bibinfo {volume} {11}},\ \bibinfo
  {pages} {362 } (\bibinfo {year} {1983})}\BibitemShut {NoStop}%
\bibitem [{\citenamefont {Hahn}\ and\ \citenamefont
  {K{\"a}rger}(1996)}]{Hahn/Kaerger:1996}%
  \BibitemOpen
  \bibfield  {author} {\bibinfo {author} {\bibfnamefont {K.}~\bibnamefont
  {Hahn}}\ and\ \bibinfo {author} {\bibfnamefont {J.}~\bibnamefont
  {K{\"a}rger}},\ }\bibfield  {title} {\bibinfo {title} {Molecular dynamics
  simulation of single-file systems},\ }\href
  {https://doi.org/10.1021/jp951807u} {\bibfield  {journal} {\bibinfo
  {journal} {J. Phys. Chem.}\ }\textbf {\bibinfo {volume} {100}},\ \bibinfo
  {pages} {316} (\bibinfo {year} {1996})}\BibitemShut {NoStop}%
\bibitem [{\citenamefont {Kollmann}(2003)}]{Kollmann:2003}%
  \BibitemOpen
  \bibfield  {author} {\bibinfo {author} {\bibfnamefont {M.}~\bibnamefont
  {Kollmann}},\ }\bibfield  {title} {\bibinfo {title} {Single-file diffusion of
  atomic and colloidal systems: Asymptotic laws},\ }\href
  {https://doi.org/10.1103/PhysRevLett.90.180602} {\bibfield  {journal}
  {\bibinfo  {journal} {Phys. Rev. Lett.}\ }\textbf {\bibinfo {volume} {90}},\
  \bibinfo {pages} {180602} (\bibinfo {year} {2003})}\BibitemShut {NoStop}%
\bibitem [{\citenamefont {Lizana}\ and\ \citenamefont
  {Ambj\"ornsson}(2008)}]{Lizana/Ambjornsson:2008}%
  \BibitemOpen
  \bibfield  {author} {\bibinfo {author} {\bibfnamefont {L.}~\bibnamefont
  {Lizana}}\ and\ \bibinfo {author} {\bibfnamefont {T.}~\bibnamefont
  {Ambj\"ornsson}},\ }\bibfield  {title} {\bibinfo {title} {Single-file
  diffusion in a box},\ }\href {https://doi.org/10.1103/PhysRevLett.100.200601}
  {\bibfield  {journal} {\bibinfo  {journal} {Phys. Rev. Lett.}\ }\textbf
  {\bibinfo {volume} {100}},\ \bibinfo {pages} {200601} (\bibinfo {year}
  {2008})}\BibitemShut {NoStop}%
\bibitem [{\citenamefont {Dvoyashkin}\ \emph {et~al.}(2013)\citenamefont
  {Dvoyashkin}, \citenamefont {Wang}, \citenamefont {Vasenkov},\ and\
  \citenamefont {Bowers}}]{Dvoyashkin/etal:2013}%
  \BibitemOpen
  \bibfield  {author} {\bibinfo {author} {\bibfnamefont {M.}~\bibnamefont
  {Dvoyashkin}}, \bibinfo {author} {\bibfnamefont {A.}~\bibnamefont {Wang}},
  \bibinfo {author} {\bibfnamefont {S.}~\bibnamefont {Vasenkov}},\ and\
  \bibinfo {author} {\bibfnamefont {C.~R.}\ \bibnamefont {Bowers}},\ }\bibfield
   {title} {\bibinfo {title} {Xenon in l-alanyl-l-valine nanochannels: A highly
  ideal molecular single-file system},\ }\href
  {https://doi.org/10.1021/jz4016712} {\bibfield  {journal} {\bibinfo
  {journal} {J. Phys. Chem. Lett.}\ }\textbf {\bibinfo {volume} {4}},\ \bibinfo
  {pages} {3263} (\bibinfo {year} {2013})}\BibitemShut {NoStop}%
\bibitem [{\citenamefont {Ryabov}(2016)}]{Ryabov:2016}%
  \BibitemOpen
  \bibfield  {author} {\bibinfo {author} {\bibfnamefont {A.}~\bibnamefont
  {Ryabov}},\ }\href {https://doi.org/10.1007/978-3-319-27188-0} {\emph
  {\bibinfo {title} {Stochastic Dynamics and Energetics of Biomolecular
  Systems}}},\ Springer Theses\ (\bibinfo  {publisher} {Springer, Cham},\
  \bibinfo {year} {2016})\BibitemShut {NoStop}%
\bibitem [{\citenamefont {Dolai}\ \emph {et~al.}(2020)\citenamefont {Dolai},
  \citenamefont {Das}, \citenamefont {Kundu}, \citenamefont {Dasgupta},
  \citenamefont {Dhar},\ and\ \citenamefont {Kumar}}]{Dolai/etal:2020}%
  \BibitemOpen
  \bibfield  {author} {\bibinfo {author} {\bibfnamefont {P.}~\bibnamefont
  {Dolai}}, \bibinfo {author} {\bibfnamefont {A.}~\bibnamefont {Das}}, \bibinfo
  {author} {\bibfnamefont {A.}~\bibnamefont {Kundu}}, \bibinfo {author}
  {\bibfnamefont {C.}~\bibnamefont {Dasgupta}}, \bibinfo {author}
  {\bibfnamefont {A.}~\bibnamefont {Dhar}},\ and\ \bibinfo {author}
  {\bibfnamefont {K.~V.}\ \bibnamefont {Kumar}},\ }\bibfield  {title} {\bibinfo
  {title} {Universal scaling in active single-file dynamics},\ }\href
  {https://doi.org/10.1039/D0SM00687D} {\bibfield  {journal} {\bibinfo
  {journal} {Soft Matter}\ }\textbf {\bibinfo {volume} {16}},\ \bibinfo {pages}
  {7077} (\bibinfo {year} {2020})}\BibitemShut {NoStop}%
\bibitem [{\citenamefont {Wittmann}\ \emph {et~al.}(2021)\citenamefont
  {Wittmann}, \citenamefont {L\"owen},\ and\ \citenamefont
  {Brader}}]{Wittmann/etal:2021}%
  \BibitemOpen
  \bibfield  {author} {\bibinfo {author} {\bibfnamefont {R.}~\bibnamefont
  {Wittmann}}, \bibinfo {author} {\bibfnamefont {H.}~\bibnamefont {L\"owen}},\
  and\ \bibinfo {author} {\bibfnamefont {J.~M.}\ \bibnamefont {Brader}},\
  }\bibfield  {title} {\bibinfo {title} {Order-preserving dynamics in one
  dimension -- single-file diffusion and caging from the perspective of
  dynamical density functional theory},\ }\href
  {https://doi.org/10.1080/00268976.2020.1867250} {\bibfield  {journal}
  {\bibinfo  {journal} {Mol. Phys.}\ }\textbf {\bibinfo {volume} {119}},\
  \bibinfo {pages} {e1867250} (\bibinfo {year} {2021})}\BibitemShut {NoStop}%
\bibitem [{\citenamefont {Banerjee}\ \emph {et~al.}(2022)\citenamefont
  {Banerjee}, \citenamefont {Jack},\ and\ \citenamefont
  {Cates}}]{Banerjee/etal:2022}%
  \BibitemOpen
  \bibfield  {author} {\bibinfo {author} {\bibfnamefont {T.}~\bibnamefont
  {Banerjee}}, \bibinfo {author} {\bibfnamefont {R.~L.}\ \bibnamefont {Jack}},\
  and\ \bibinfo {author} {\bibfnamefont {M.~E.}\ \bibnamefont {Cates}},\
  }\bibfield  {title} {\bibinfo {title} {Role of initial conditions in
  one-dimensional diffusive systems: Compressibility, hyperuniformity, and
  long-term memory},\ }\href {https://doi.org/10.1103/PhysRevE.106.L062101}
  {\bibfield  {journal} {\bibinfo  {journal} {Phys. Rev. E}\ }\textbf {\bibinfo
  {volume} {106}},\ \bibinfo {pages} {L062101} (\bibinfo {year}
  {2022})}\BibitemShut {NoStop}%
\bibitem [{\citenamefont {van Beijeren}\ \emph {et~al.}(1983)\citenamefont {van
  Beijeren}, \citenamefont {Kehr},\ and\ \citenamefont
  {Kutner}}]{vanBeijeren/etal:1983}%
  \BibitemOpen
  \bibfield  {author} {\bibinfo {author} {\bibfnamefont {H.}~\bibnamefont {van
  Beijeren}}, \bibinfo {author} {\bibfnamefont {K.~W.}\ \bibnamefont {Kehr}},\
  and\ \bibinfo {author} {\bibfnamefont {R.}~\bibnamefont {Kutner}},\
  }\bibfield  {title} {\bibinfo {title} {Diffusion in concentrated lattice
  gases. {III}. {T}racer diffusion on a one-dimensional lattice},\ }\href
  {https://doi.org/10.1103/PhysRevB.28.5711} {\bibfield  {journal} {\bibinfo
  {journal} {Phys. Rev. B}\ }\textbf {\bibinfo {volume} {28}},\ \bibinfo
  {pages} {5711} (\bibinfo {year} {1983})}\BibitemShut {NoStop}%
\bibitem [{\citenamefont {Leibovich}\ and\ \citenamefont
  {Barkai}(2013)}]{Leibovich/Barkai:2013}%
  \BibitemOpen
  \bibfield  {author} {\bibinfo {author} {\bibfnamefont {N.}~\bibnamefont
  {Leibovich}}\ and\ \bibinfo {author} {\bibfnamefont {E.}~\bibnamefont
  {Barkai}},\ }\bibfield  {title} {\bibinfo {title} {Everlasting effect of
  initial conditions on single-file diffusion},\ }\href
  {https://doi.org/10.1103/PhysRevE.88.032107} {\bibfield  {journal} {\bibinfo
  {journal} {Phys. Rev. E}\ }\textbf {\bibinfo {volume} {88}},\ \bibinfo
  {pages} {032107} (\bibinfo {year} {2013})}\BibitemShut {NoStop}%
\bibitem [{\citenamefont {Baxter}(1968)}]{Baxter:1968}%
  \BibitemOpen
  \bibfield  {author} {\bibinfo {author} {\bibfnamefont {R.~J.}\ \bibnamefont
  {Baxter}},\ }\bibfield  {title} {\bibinfo {title} {Percus--{Y}evick equation
  for hard spheres with surface adhesion},\ }\href
  {https://doi.org/10.1063/1.1670482} {\bibfield  {journal} {\bibinfo
  {journal} {J. Chem. Phys.}\ }\textbf {\bibinfo {volume} {49}},\ \bibinfo
  {pages} {2770} (\bibinfo {year} {1968})}\BibitemShut {NoStop}%
\bibitem [{\citenamefont {Percus}(1982)}]{Percus:1982}%
  \BibitemOpen
  \bibfield  {author} {\bibinfo {author} {\bibfnamefont {J.~K.}\ \bibnamefont
  {Percus}},\ }\bibfield  {title} {\bibinfo {title} {One-dimensional classical
  fluid with nearest-neighbor interaction in arbitrary external field},\ }\href
  {https://doi.org/10.1007/BF01011623} {\bibfield  {journal} {\bibinfo
  {journal} {J. Stat. Phys.}\ }\textbf {\bibinfo {volume} {28}},\ \bibinfo
  {pages} {67} (\bibinfo {year} {1982})}\BibitemShut {NoStop}%
\bibitem [{\citenamefont {Delfau}\ \emph {et~al.}(2011)\citenamefont {Delfau},
  \citenamefont {Coste},\ and\ \citenamefont {Saint~Jean}}]{Delfau/etal:2011}%
  \BibitemOpen
  \bibfield  {author} {\bibinfo {author} {\bibfnamefont {J.-B.}\ \bibnamefont
  {Delfau}}, \bibinfo {author} {\bibfnamefont {C.}~\bibnamefont {Coste}},\ and\
  \bibinfo {author} {\bibfnamefont {M.}~\bibnamefont {Saint~Jean}},\ }\bibfield
   {title} {\bibinfo {title} {Single-file diffusion of particles with
  long-range interactions: Damping and finite-size effects},\ }\href
  {https://doi.org/10.1103/PhysRevE.84.011101} {\bibfield  {journal} {\bibinfo
  {journal} {Phys. Rev. E}\ }\textbf {\bibinfo {volume} {84}},\ \bibinfo
  {pages} {011101} (\bibinfo {year} {2011})}\BibitemShut {NoStop}%
\bibitem [{\citenamefont {Centres}\ and\ \citenamefont
  {Bustingorry}(2010)}]{Centres/Bustingorry:2010}%
  \BibitemOpen
  \bibfield  {author} {\bibinfo {author} {\bibfnamefont {P.~M.}\ \bibnamefont
  {Centres}}\ and\ \bibinfo {author} {\bibfnamefont {S.}~\bibnamefont
  {Bustingorry}},\ }\bibfield  {title} {\bibinfo {title} {Effective
  edwards-wilkinson equation for single-file diffusion},\ }\href
  {https://doi.org/10.1103/PhysRevE.81.061101} {\bibfield  {journal} {\bibinfo
  {journal} {Phys. Rev. E}\ }\textbf {\bibinfo {volume} {81}},\ \bibinfo
  {pages} {061101} (\bibinfo {year} {2010})}\BibitemShut {NoStop}%
\bibitem [{\citenamefont {Fouad}\ and\ \citenamefont
  {Noel}(2017)}]{Fouad/Noel:2017}%
  \BibitemOpen
  \bibfield  {author} {\bibinfo {author} {\bibfnamefont {A.~M.}\ \bibnamefont
  {Fouad}}\ and\ \bibinfo {author} {\bibfnamefont {J.~A.}\ \bibnamefont
  {Noel}},\ }\bibfield  {title} {\bibinfo {title} {On the role of adhesion in
  single-file dynamics},\ }\href
  {https://doi.org/https://doi.org/10.1016/j.physa.2017.03.030} {\bibfield
  {journal} {\bibinfo  {journal} {Physica A}\ }\textbf {\bibinfo {volume}
  {480}},\ \bibinfo {pages} {1} (\bibinfo {year} {2017})}\BibitemShut {NoStop}%
\bibitem [{\citenamefont {Antonov}\ \emph {et~al.}(2022)\citenamefont
  {Antonov}, \citenamefont {Schweers}, \citenamefont {Ryabov},\ and\
  \citenamefont {Maass}}]{Antonov/etal:2022c}%
  \BibitemOpen
  \bibfield  {author} {\bibinfo {author} {\bibfnamefont {A.~P.}\ \bibnamefont
  {Antonov}}, \bibinfo {author} {\bibfnamefont {S.}~\bibnamefont {Schweers}},
  \bibinfo {author} {\bibfnamefont {A.}~\bibnamefont {Ryabov}},\ and\ \bibinfo
  {author} {\bibfnamefont {P.}~\bibnamefont {Maass}},\ }\bibfield  {title}
  {\bibinfo {title} {Brownian dynamics simulations of hard rods in external
  fields and with contact interactions},\ }\href
  {https://doi.org/10.1103/PhysRevE.106.054606} {\bibfield  {journal} {\bibinfo
   {journal} {Phys. Rev. E}\ }\textbf {\bibinfo {volume} {106}},\ \bibinfo
  {pages} {054606} (\bibinfo {year} {2022})}\BibitemShut {NoStop}%
\bibitem [{\citenamefont {Miller}\ and\ \citenamefont
  {Frenkel}(2004{\natexlab{a}})}]{Miller/Frenkel:2004_1}%
  \BibitemOpen
  \bibfield  {author} {\bibinfo {author} {\bibfnamefont {M.~A.}\ \bibnamefont
  {Miller}}\ and\ \bibinfo {author} {\bibfnamefont {D.}~\bibnamefont
  {Frenkel}},\ }\bibfield  {title} {\bibinfo {title} {Simulating colloids with
  {Baxter's} adhesive hard sphere model},\ }\href
  {https://doi.org/10.1088/0953-8984/16/42/008} {\bibfield  {journal} {\bibinfo
   {journal} {J. Phys.: Condens. Matter}\ }\textbf {\bibinfo {volume} {16}},\
  \bibinfo {pages} {S4901} (\bibinfo {year} {2004}{\natexlab{a}})}\BibitemShut
  {NoStop}%
\bibitem [{\citenamefont {Miller}\ and\ \citenamefont
  {Frenkel}(2004{\natexlab{b}})}]{Miller/Frenkel:2004_2}%
  \BibitemOpen
  \bibfield  {author} {\bibinfo {author} {\bibfnamefont {M.~A.}\ \bibnamefont
  {Miller}}\ and\ \bibinfo {author} {\bibfnamefont {D.}~\bibnamefont
  {Frenkel}},\ }\bibfield  {title} {\bibinfo {title} {Phase diagram of the
  adhesive hard sphere fluid},\ }\href {https://doi.org/10.1063/1.1758693}
  {\bibfield  {journal} {\bibinfo  {journal} {J. Chem. Phys.}\ }\textbf
  {\bibinfo {volume} {121}},\ \bibinfo {pages} {535} (\bibinfo {year}
  {2004}{\natexlab{b}})}\BibitemShut {NoStop}%
\bibitem [{Note1()}]{Note1}%
  \BibitemOpen
  \bibinfo {note} {The simulations have been performed under periodic boundary
  conditions for typical system sizes of $800\sigma $. To obtain good numerical
  accuracy, averages were taken over 250 equilibrated initial particle
  configurations. At stickiness $\gamma =10$, the CPU time for a simulation run
  of one configuration was 5h for $\rho \sigma =0.25$ and 45h for $\rho \sigma
  =0.8$ on one core of an AMD EPYC 7742 processor.}\BibitemShut {Stop}%
\bibitem [{\citenamefont {Salsburg}\ \emph {et~al.}(1953)\citenamefont
  {Salsburg}, \citenamefont {Zwanzig},\ and\ \citenamefont
  {Kirkwood}}]{Salsburg/etal:1953}%
  \BibitemOpen
  \bibfield  {author} {\bibinfo {author} {\bibfnamefont {Z.~W.}\ \bibnamefont
  {Salsburg}}, \bibinfo {author} {\bibfnamefont {R.~W.}\ \bibnamefont
  {Zwanzig}},\ and\ \bibinfo {author} {\bibfnamefont {J.~G.}\ \bibnamefont
  {Kirkwood}},\ }\bibfield  {title} {\bibinfo {title} {Molecular distribution
  functions in a one‐dimensional fluid},\ }\href
  {https://doi.org/10.1063/1.1699116} {\bibfield  {journal} {\bibinfo
  {journal} {J. Chem. Phys.}\ }\textbf {\bibinfo {volume} {21}},\ \bibinfo
  {pages} {1098} (\bibinfo {year} {1953})}\BibitemShut {NoStop}%
\bibitem [{\citenamefont {Tago}\ and\ \citenamefont
  {Katsura}(1975)}]{Tago/Katsura:1975}%
  \BibitemOpen
  \bibfield  {author} {\bibinfo {author} {\bibfnamefont {Y.}~\bibnamefont
  {Tago}}\ and\ \bibinfo {author} {\bibfnamefont {S.}~\bibnamefont {Katsura}},\
  }\bibfield  {title} {\bibinfo {title} {The {P}ercus--{Y}evick equation of
  state for the sticky hard rod gas},\ }\href {https://doi.org/10.1139/p75-314}
  {\bibfield  {journal} {\bibinfo  {journal} {Can. J. Phys.}\ }\textbf
  {\bibinfo {volume} {53}},\ \bibinfo {pages} {2587} (\bibinfo {year}
  {1975})}\BibitemShut {NoStop}%
\bibitem [{\citenamefont {Yuste}\ and\ \citenamefont
  {Santos}(1993)}]{Yuste/Santos:1993}%
  \BibitemOpen
  \bibfield  {author} {\bibinfo {author} {\bibfnamefont {S.~B.}\ \bibnamefont
  {Yuste}}\ and\ \bibinfo {author} {\bibfnamefont {A.}~\bibnamefont {Santos}},\
  }\bibfield  {title} {\bibinfo {title} {Radial distribution function for
  sticky hard-core fluids},\ }\href {https://doi.org/10.1007/BF01048029}
  {\bibfield  {journal} {\bibinfo  {journal} {J. Stat. Phys.}\ }\textbf
  {\bibinfo {volume} {72}},\ \bibinfo {pages} {703} (\bibinfo {year}
  {1993})}\BibitemShut {NoStop}%
\bibitem [{\citenamefont {Misiunas}\ and\ \citenamefont
  {Keyser}(2019)}]{Misiunas/Keyser:2019}%
  \BibitemOpen
  \bibfield  {author} {\bibinfo {author} {\bibfnamefont {K.}~\bibnamefont
  {Misiunas}}\ and\ \bibinfo {author} {\bibfnamefont {U.~F.}\ \bibnamefont
  {Keyser}},\ }\bibfield  {title} {\bibinfo {title} {Density-dependent speed-up
  of particle transport in channels},\ }\href
  {https://doi.org/10.1103/PhysRevLett.122.214501} {\bibfield  {journal}
  {\bibinfo  {journal} {Phys. Rev. Lett.}\ }\textbf {\bibinfo {volume} {122}},\
  \bibinfo {pages} {214501} (\bibinfo {year} {2019})}\BibitemShut {NoStop}%
\bibitem [{\citenamefont {Driscoll}\ and\ \citenamefont
  {Delmotte}(2019)}]{Driscoll/Delmotte:2019}%
  \BibitemOpen
  \bibfield  {author} {\bibinfo {author} {\bibfnamefont {M.}~\bibnamefont
  {Driscoll}}\ and\ \bibinfo {author} {\bibfnamefont {B.}~\bibnamefont
  {Delmotte}},\ }\bibfield  {title} {\bibinfo {title} {Leveraging collective
  effects in externally driven colloidal suspensions: {E}xperiments and
  simulations},\ }\href
  {https://doi.org/https://doi.org/10.1016/j.cocis.2018.10.002} {\bibfield
  {journal} {\bibinfo  {journal} {Curr. Opin. Colloid Interface Sci.}\ }\textbf
  {\bibinfo {volume} {40}},\ \bibinfo {pages} {42} (\bibinfo {year}
  {2019})}\BibitemShut {NoStop}%
\end{thebibliography}

\begin{thebibliography}{3}%
\makeatletter
\providecommand \@ifxundefined [1]{%
 \@ifx{#1\undefined}
}%
\providecommand \@ifnum [1]{%
 \ifnum #1\expandafter \@firstoftwo
 \else \expandafter \@secondoftwo
 \fi
}%
\providecommand \@ifx [1]{%
 \ifx #1\expandafter \@firstoftwo
 \else \expandafter \@secondoftwo
 \fi
}%
\providecommand \natexlab [1]{#1}%
\providecommand \enquote  [1]{``#1''}%
\providecommand \bibnamefont  [1]{#1}%
\providecommand \bibfnamefont [1]{#1}%
\providecommand \citenamefont [1]{#1}%
\providecommand \href@noop [0]{\@secondoftwo}%
\providecommand \href [0]{\begingroup \@sanitize@url \@href}%
\providecommand \@href[1]{\@@startlink{#1}\@@href}%
\providecommand \@@href[1]{\endgroup#1\@@endlink}%
\providecommand \@sanitize@url [0]{\catcode `\\12\catcode `\$12\catcode
  `\&12\catcode `\#12\catcode `\^12\catcode `\_12\catcode `\%12\relax}%
\providecommand \@@startlink[1]{}%
\providecommand \@@endlink[0]{}%
\providecommand \url  [0]{\begingroup\@sanitize@url \@url }%
\providecommand \@url [1]{\endgroup\@href {#1}{\urlprefix }}%
\providecommand \urlprefix  [0]{URL }%
\providecommand \Eprint [0]{\href }%
\providecommand \doibase [0]{https://doi.org/}%
\providecommand \selectlanguage [0]{\@gobble}%
\providecommand \bibinfo  [0]{\@secondoftwo}%
\providecommand \bibfield  [0]{\@secondoftwo}%
\providecommand \translation [1]{[#1]}%
\providecommand \BibitemOpen [0]{}%
\providecommand \bibitemStop [0]{}%
\providecommand \bibitemNoStop [0]{.\EOS\space}%
\providecommand \EOS [0]{\spacefactor3000\relax}%
\providecommand \BibitemShut  [1]{\csname bibitem#1\endcsname}%
\let\auto@bib@innerbib\@empty
\bibitem [{\citenamefont {Kollmann}(2003)}]{Kollmann:2003}%
  \BibitemOpen
  \bibfield  {author} {\bibinfo {author} {\bibfnamefont {M.}~\bibnamefont
  {Kollmann}},\ }\bibfield  {title} {\bibinfo {title} {Single-file diffusion of
  atomic and colloidal systems: Asymptotic laws},\ }\href
  {https://doi.org/10.1103/PhysRevLett.90.180602} {\bibfield  {journal}
  {\bibinfo  {journal} {Phys. Rev. Lett.}\ }\textbf {\bibinfo {volume} {90}},\
  \bibinfo {pages} {180602} (\bibinfo {year} {2003})}\BibitemShut {NoStop}%
\bibitem [{\citenamefont {N{\"a}gele}(1996)}]{Naegele:1996}%
  \BibitemOpen
  \bibfield  {author} {\bibinfo {author} {\bibfnamefont {G.}~\bibnamefont
  {N{\"a}gele}},\ }\bibfield  {title} {\bibinfo {title} {On the dynamics and
  structure of charge-stabilized suspensions},\ }\href
  {https://doi.org/https://doi.org/10.1016/0370-1573(95)00078-X} {\bibfield
  {journal} {\bibinfo  {journal} {Phys. Rep.}\ }\textbf {\bibinfo {volume}
  {272}},\ \bibinfo {pages} {215} (\bibinfo {year} {1996})}\BibitemShut
  {NoStop}%
\bibitem [{\citenamefont {Yuste}\ and\ \citenamefont
  {Santos}(1993)}]{Yuste/Santos:1993}%
  \BibitemOpen
  \bibfield  {author} {\bibinfo {author} {\bibfnamefont {S.~B.}\ \bibnamefont
  {Yuste}}\ and\ \bibinfo {author} {\bibfnamefont {A.}~\bibnamefont {Santos}},\
  }\bibfield  {title} {\bibinfo {title} {Radial distribution function for
  sticky hard-core fluids},\ }\href {https://doi.org/10.1007/BF01048029}
  {\bibfield  {journal} {\bibinfo  {journal} {J. Stat. Phys.}\ }\textbf
  {\bibinfo {volume} {72}},\ \bibinfo {pages} {703} (\bibinfo {year}
  {1993})}\BibitemShut {NoStop}%
\end{thebibliography}

%

\end{document}